\documentclass[journal=jpcafh,manuscript=article,layout=twocolumn]{achemso}
\SectionNumbersOn
\newcommand{\onlinecite}[1]{\hspace{-1 ex} \nocite{#1}\citenum{#1}} 

\usepackage[utf8]{inputenc}

\usepackage{graphicx} 
\usepackage[dvipsnames]{xcolor}

\usepackage{amsmath}
\usepackage{amssymb,latexsym}
\usepackage{braket}
\usepackage{appendix}
\usepackage{comment}

\usepackage{tikz}
\usetikzlibrary{cd}

\renewcommand{\Re}{\operatorname{Re}}
\renewcommand{\Im}{\operatorname{Im}}
\newcommand{\Tr}{\operatorname{Tr}}
\newcommand{\RR}{\mathbb{R}}
\newcommand{\CC}{\mathbb{C}}
\newcommand{\HH}{\mathbb{H}}

\newcommand{\calA}{\mathcal{A}}
\newcommand{\calH}{\mathcal{H}}
\newcommand{\calE}{\mathcal{E}}

\newcommand{\calC}{\mathcal{C}}

\newcommand{\calL}{\mathcal{L}}

\newcommand{\calV}{\mathcal{V}}
\newcommand{\calM}{\mathcal{M}}
\newcommand{\calN}{\mathcal{N}}

\DeclareMathOperator*{\argmin}{argmin}

\newcommand{\llangle}{\langle\!\langle}
\newcommand{\rrangle}{\rangle\!\rangle}
\newcommand{\dualpairing}[1]{\llangle{#1}\rrangle}

\title{The time-dependent bivariational principle: Theoretical foundation for real-time propagation methods of coupled-cluster type}
\author{Simen Kvaal}
\email{simen.kvaal@kjemi.uio.no}
\affiliation{Hylleraas Centre for Quantum Molecular Sciences, Department of Chemistry, University of Oslo, P.O. Box 1033 Blindern, N-0315 Oslo, Norway 
}%
\author{Håkon Fredheim}
\affiliation{Hylleraas Centre for Quantum Molecular Sciences, Department of Chemistry, University of Oslo, P.O. Box 1033 Blindern, N-0315 Oslo, Norway 
}%
\author{Mads Greisen Højlund}
\affiliation{Department of Chemistry, Aarhus University, Langelandsgade 140, 8000 Aarhus C, Denmark}
\affiliation{Hylleraas Centre for Quantum Molecular Sciences, Department of Chemistry, University of Oslo, P.O. Box 1033 Blindern, N-0315 Oslo, Norway 
}%
\author{Thomas Bondo Pedersen}
\affiliation{Hylleraas Centre for Quantum Molecular Sciences, Department of Chemistry, University of Oslo, P.O. Box 1033 Blindern, N-0315 Oslo, Norway 
}%

\date{\today}

\begin{document}

\begin{abstract}
Real-time propagation methods for chemistry and physics are invariably formulated using variational techniques. The time-dependent bivariational principle (TD-BIVP) is known to be the proper framework for coupled-cluster type methods, and is here studied from a differential geometric point of view. It is demonstrated how two distinct classical Hamilton's equations of motion arise from considering the real and imaginary parts of the action integral. The latter is new, and can in principle be used to develop novel propagation methods. Conservation laws and Poisson brackets are introduced, completing the analogy with classical mechanics. An overview of established real-time propagation methods is given in the context of our formulation of the TD-BIVP, namely time-dependent traditional coupled-cluster theory, orbital-adaptive coupled-cluster theory, time-dependent orthogonal optimized coupled-cluster theory, and equation-of-motion coupled cluster theory.
\end{abstract}

\maketitle

\section{Introduction}
\label{sec:introduction}

With the increase in demand for high-accuracy first-principles simulations of the quantum dynamics of molecular systems at the attosecond time scale comes the need for theoretical frameworks that allow derivation of affordable, accurate, and systematically improvable computational tools. The cornerstone of approximate quantum dynamics has for more than half a century been the \emph{time-dependent variational principle} (TDVP) and the \emph{McLachlan variational principle}, often collecively reffered to as the Dirac--Frenkel--McLachlan variational principle, even though they are not equivalent in all circumstances.\cite{dirac_principles_1958,frenkel_wave_1934,mclachlan_variational_1964,kramer_geometry_1981,lubich_quantum_2008,yuan_theory_2019,feldmeier_molecular_2000,deumens_time-dependent_1994}. However, the most popular wavefunction-based method in quantum chemistry is the coupled-cluster (CC) method, which is notable for being \emph{not variational} -- it is \emph{bivariational}. 

In this article, we present a comprehensive study of \emph{the time-dependent bivariational principle} (TD-BIVP)\cite{chernoff_properties_1974,arponen_variational_1983} which generalizes the TDVP, and forms the proper setting for the various forms of time-dependent CC theory that have been developed over the last decades; from time-dependent traditional CC theory\cite{pedersen_symplectic_2019,peyton_reduced_2023}, via equation-of-motion CC theory,\cite{nascimento_linear_2016,nascimento_general_2019} to orbital-adaptive and orbital-optimized CC theory.\cite{kvaal_ab_2012,madsen_time-dependent_2020} Applications of time-dependent CC theory with a bivariational formulation covers applications as diverse as electronic-structure theory\cite{li_real-time_2020,pedersen_symplectic_2019}, the vibrational Schrödinger equation\cite{hansen_time-dependent_2019,hansen_extended_2020,madsen_general_2020,madsen_time-dependent_2020}, and nuclear structure theory\cite{pigg_time-dependent_2012}. For a recent review, see Ref.~\onlinecite{sverdrup_ofstad_time-dependent_2023}. Considering the importance of variational principles for the development of computational tools on one hand, and the growing importance of time-dependent CC theory on the other, establishing the theoretical framework of the TD-BIVP can therefore catalyze rapid progress in the development of sophisticated and relatively low-cost real-time propagation methods in several fields, and in particular in attochemistry.

The TD-BIVP was first mentioned in passing by Chernoff and Marsden\cite{chernoff_properties_1974}, without coining the term, who devised a Lagrangian density (in the sense of field theory), and a corresponding action $\calA$, in which the system wavefunction and its complex conjugate were formally independent variables $\psi$ and $\tilde{\psi}$ forming canonically conjugate variables in an abstract phase space $\HH$. The Euler--Lagrange equations were the time-dependent Schrödinger equation and its dual, written as a pair of Hamilton's equations of motion on complex form. The principle was independently discovered by Arponen in his seminal treatise on coupled-cluster (CC) theory\cite{arponen_variational_1983}, where the CC amplitudes turn out to be canonical variables, preserving the form of Hamilton's equations of motion. In later publications by the trio of Arponen, Bishop and Pajanne on the extended CC method, the principle was occasionally invoked, emphasizing the canonical structure of these methods, including small oscillations around the ground state solution\cite{arponen_extended_1987,arponen_extended_1987-1,arponen_independent-cluster_1991,arponen_independent-cluster_1993,arponen_independent-cluster_1993-1}. A more formal symplectic geometry formulation of coupled-cluster theory was first considered by Arponen\cite{arponen_constrained_1997}, who identified a real Hamiltonian system; see Sec.~\ref{sec:real-ham} in this article.

While the canonical transformation of time-dependent CC theory is an exact reformulation of quantum dynamics, approximations are invariably introduced in the form of the conventional hierarchy of CC with singles, doubles, triples, etc. Such approximations amount to choosing a submanifold $\calM \subset \HH$ of phase space and restricting variations of the variables in the action $\calA$ to be in the tangent space of $\calM$, in a similar fashion as is done for the Dirac--Frenkel and McLachlan principles.\cite{lubich_quantum_2008} Traditionally, the manifold $\calM$ has been assumed to be complex in the bivariational case\cite{kvaal_ab_2012,kvaal_variational_2013,hansen_time-dependent_2019,pedersen_symplectic_2019}, i.e., the local coordinates are complex numbers and the points on $\calM$ are complex differentiable with respect to the coordinates. The argument has been that since $\calA$ is complex-valued, the manifold must be complex in order to give well-defined equations of motion. In this article, we show that this can be refined, and a careful consideration of real manifolds yields two distinct time-dependent bivariational bivariational principles, $\delta \Re \calA = 0$ and $\delta\Im\calA = 0$, that, just like the Dirac--Frenkel and McLachlan principles, are equivalent when the tangent spaces of $\calM$ are complex vector spaces, and exact when $\calM=\HH$, the full phase space. For general real manifolds, the variational principles are distinct. The principle $\delta \Re\calA=0$ is exemplified by the time-dependent orbital-optimized CC method by Sato and coworkers\cite{satoCommunicationTimedependentOptimized2018}, while the second variational principle has not been put to use, to our knowledge.

The remainder of this article is structured as follows.  In Section~\ref{sec:bivp} we introduce the TD-BIVP, with emphasis on symplectic geometry. In Section~\ref{sec:manifold-evolution} we consider the restriction of the bivariational dynamics to (symplectic) submanifolds, and develop the equations of motions in local coordinates. We consider both real and complex submanifolds. We also discuss the relation of the TD-BIVP to the Dirac--Frenkel and McLachlan variational principles. In Section~\ref{sec:current-methods}, we formulate bivariational dynamics methods in the literature using the present abstract framework. Finally, in Section~\ref{sec:conclusion} we present our conclusion and future perspectives. An appendix provides additional details.

\section{The time-dependent bivariational principle}
\label{sec:bivp}

The time-dependent bivariational principle is a stationary-action principle for the time-dependent Schrödinger equation and its dual. Consider the action-like functional
\begin{equation}
    \calA = \int_0^T i\braket{\tilde{\psi}|\dot{\psi}} -  \braket{\tilde{\psi}|H|\psi} \; dt. \label{eq:action}
\end{equation}
We assume for simplicity that the system Hamiltonian $H$ is a bounded self-adjoint operator, in order to avoid unnecessary formal complexity. Here, $\psi(t) \in \calH$, a wavefunction in a complex (separable) Hilbert space, and where $\tilde{\psi}(t) \in \calH^*$, the complex-conjugate Hilbert space, or dual space. The notation $\braket{\cdot|\cdot}$ is thus the dual pairing on $\calH^*\times\calH$. (Note that $\tilde{\psi}$ is \emph{not} meant to be the complex conjugate of $\psi$, but an independent variable that lives in dual space.) In physics terms, $\calH$ is the space of ``kets'', while $\calH^*$ is the space of ``bras''. 

The action is stationary ($\delta\calA = 0$) under arbitrary smooth variations (vanishing at the endpoints) of $\psi$ and $\tilde{\psi}$ if and only if
\begin{equation}
     i\dot{\psi} = H\psi, \quad -i\dot{\tilde{\psi}}  = H^t \tilde{\psi},
    \label{eq:tdse-and-dual}
\end{equation}
i.e., the action principle is equivalent to the time-dependent Schrödinger equation and its dual. Here, $H^t \tilde{\psi}=\bra{\tilde{\psi}}H$ in bra--ket notation is called the Banach adjoint, or operator transpose\cite{lionsMathematicalAnalysisNumerical1988}. With the Hamiltonian function $\calE(\tilde{\psi},\psi)=\braket{\tilde{\psi}|H \psi}$, Eq.~\eqref{eq:tdse-and-dual} takes the form of a complex set of Hamilton's equations of motion,
\begin{equation}
    i \dot{\psi} = \frac{\partial\calE}{\partial\tilde{\psi}}, \quad\text{and}\quad -i\dot{\tilde{\psi}} = \frac{\partial\calE}{\partial \psi} ,
    \label{eq:complex-hamiltonian-system}
\end{equation}
which comes as no surprise when $\delta\calA=0$ is recognized as the \emph{Modified Hamilton's Principle} from classical mechanics~\cite{goldstein_classical_2008}.  Thus, $\HH \equiv \calH^*\oplus \calH$ serves as phase space, with $(\tilde{\psi},\psi)\in\HH$ forming a pair of (infinite-dimensional) momenta and coordinate vectors, respectively.

The action $\cal{A}$ implies two conserved quantities: First, the overlap is conserved, $d\braket{\tilde{\psi}|\psi}/dt=0$, and second, the energy is conserved, $d\calE/dt = 0$. 

We make note of the small curiosity, also present for the Modified Hamilton's Principle for classical dynamics, that only if $(\psi(T), \tilde{\psi}(T))$ are \emph{actually} solutions of Eqs.~\eqref{eq:complex-hamiltonian-system} with initial conditions $(\tilde{\psi}(0),\psi(0))$, will $\calA$ have a critical point. On the other hand, the variations do not ``see'' the boundary conditions, since they are supported in the interior of the time interval $[0,T]$. Thus, the action functional can be viewed as a semi-local integral formulation of the time-dependent Schrödinger equation, and we will omit the specification of the time boundaries when writing the action integrals.

In exact quantum mechanics and in the TDVP, the only allowed initial conditions satisfy $\tilde{\psi}(0) = \psi(0)^\dag/\|\psi(0)\|^2$, and consequently $\tilde{\psi}(t) = \psi(t)^\dag/\|\psi(t)\|^2$ for all $t$.  (In general, $\psi^\dag$ is given by Riesz' representation theorem~\cite{kreyszig_introductory_1978}. If $\calH$ is a space of square-integrable functions, $\psi^\dag$ can be taken to be the complex conjugate function.) However, the power of the bivariational principle is that it allows far more flexible approximation schemes than the TDVP, since we may allow $\tilde{\psi}$ and $\psi$ to have independent approximations, such as is the case in coupled-cluster theory.

\subsection{Complex Hamiltonian systems as real Hamiltonian systems}
\label{sec:real-ham}

Complex Hamiltonian equations of motion may seem strange and very different from the standard real Hamiltonian systems of classical mechanics. In this section we demonstrate, however, that the complex Hamiltonian system is in fact just real Hamiltonian systems in disguise. 

The real and imaginary parts of the action $\calA$  must be simultaneously stationary if $\delta\calA=0$. Following the idea of Arponen \emph{et al.}\cite{arponen_constrained_1997}, we show that the complex Hamiltonian system is in fact equivalent to a standard real-valued Hamiltonian system: Let $\{\phi_\mu\}\subset\calH$ and $\{\tilde{\phi}^\mu\}\subset \calH^*$ be biorthogonal bases, i.e. $\braket{\tilde{\phi}^\mu,\phi_\nu} = \delta^\mu_\nu$, and define real-valued vectors $q_i$ and $p_i$, $i\in\{1,2\}$, such that
\begin{equation}
    \psi = \sum_\mu \phi_\mu (q_1^\mu + i p_2^\mu), 
    \quad \tilde{\psi} = \sum_\mu \tilde{\phi}^\mu (q_{2,\mu} - i p_{1,\mu}).
\end{equation}
We obtain, up to a total time derivative,
\begin{subequations}
    \begin{align}
        \Re \calA = \int p_1\cdot\dot{q}_1 + p_2\cdot\dot{q}_2 - \Re \calE \; dt, \\\intertext{and}
        \Im \calA = \int q_2\cdot \dot{q}_1 + p_1\cdot\dot{p}_2 - \Im \calE \; dt.
    \end{align}
\end{subequations}
We recognize the real part of $\calA$ as the Lagrangian from the Modified Hamilton's Principle\cite{goldstein_classical_2008}, and consequently,
\begin{subequations}
\begin{equation}
    \dot{q}_i = \frac{\partial\Re\calE}{\partial p_i}, \quad \text{and}\quad \dot{p}_i = -\frac{\partial\Re\calE}{\partial q_i}. \label{eq:re-ham-1}
\end{equation}
For the imaginary part, set $(Q_1,Q_2) = (q_1, p_2)$ and $(P_1,P_2) = (q_2, p_1)$ to obtain another Hamiltonian system
\begin{equation}
    \dot{Q}_i = \frac{\partial\Im\calE}{\partial P_i}, \quad \text{and}
\quad \dot{P}_i = -\frac{\partial\Im\calE}{\partial Q_i}. \label{eq:re-ham-2}
\end{equation}
\end{subequations}
Since $\cal{A}$ is complex differentiable, Eqs.~\eqref{eq:re-ham-1} and \eqref{eq:re-ham-2} can be seen to be related by the Cauchy--Riemann equations, and hence equivalent. We make the observation that the complex Hamiltonian system~\eqref{eq:complex-hamiltonian-system} is equivalent to two \emph{distinct} standard real Hamiltonian systems. Hamiltonian systems with multiple distinct symplectic structures are called \emph{bi-Hamiltonian systems} \cite{olver_canonical_1993}.

\subsection{Symplectic formulation of the bivariational principle}

We now develop the theory of the bivariational principle using a more abstract approach. This will in the end produce very concise and informative expressions. We equip our phase space $\HH$ with a symplectic form $\omega : \HH \times \HH \to \CC$, an antisymmetric and nondegenerate bilinear map, which we define using the expression
\begin{equation}
    \omega(u, v) = \braket{\tilde{\psi}_u| \psi_v} - \braket{\tilde{\psi}_v| \psi_u} = \dualpairing{u,Jv},
\end{equation}
where $u=(\tilde\psi_u,\psi_u)$, $v = (\tilde \psi_v,\psi_v)$, and $J(\tilde{\psi},\psi) \equiv (\psi, -\tilde{\psi}) \in \mathbb{H}^*$.
The notation $\dualpairing{u,Jv}$ is the dual pairing of $\HH = \calH^* \oplus \calH$ and $\HH^* = \calH \oplus \calH^*$, and does \emph{not} denote the Hermitian inner product on $\HH$. In fact, we will never use an inner product on $\HH$. (The reader may note that in this particular dual pairing -- natural from the context -- the dual element is to the \emph{right}, as opposed to the left as in $\braket{\cdot|\cdot}$.)   Consequently, $(\HH,\omega)$ is now a (linear) symplectic space.

For an operator $A : \calH\to\calH$, let us introduce the ``symmetrization'' $\hat{A} : \mathbb{H}\to\mathbb{H}^*$ as
\begin{equation}
    \hat{A} (\tilde{\psi}, \psi) =  \frac{1}{2}(A\psi, A^t\tilde{\psi}),
    \label{Equation: Symmetrization}
\end{equation}
and the corresponding bilinear functional $\calE(u,v) = \dualpairing{u,\hat{H}v} = (\braket{\tilde{\psi}_u|H\psi_v} + \braket{\tilde{\psi}_v|H\psi_u})/2$. We define the notation $\calE(u) \equiv \calE(u,u) = \braket{\tilde{\psi}_u|H\psi_u}$. The action functional reads, up to a total time derivative,
\begin{equation}
    \calA =  \int_0^T \frac{i}{2} \omega(u,\dot{u}) - \calE(u) \; dt. \label{eq:action-abstract}
\end{equation}
Let $\delta u(t) \in \HH$ be an arbitrary variation. Since $\omega$ is antisymmetric, and $\calE$ is symmetric, integration by parts readily yields
\begin{equation}
    \begin{split}
    \delta \calA &= \int i \omega(\delta u, \dot{u}) -  d\calE(u)(\delta u) \; dt \\
    &= \int i \dualpairing{\delta u,J\dot{u}} - 2\dualpairing{\delta u, \hat{H} u}\; dt.
    \label{eq:variation-action-symp}
    \end{split}
\end{equation}
Here, $d\calE(u)$ is the Fréchet derivative  of $\calE$, and $\delta\calE = d\calE(u)(\delta u)$ is correspondingly the directional derivative of $\calE$ in the direction $\delta u$. Since $\delta u$ was arbitrary, Hamilton's equations of motion~\eqref{eq:complex-hamiltonian-system} become
\begin{equation}
    i\dot{u} = J^{-1} d\calE(u), \label{eq:complex-hamiltonian-system-phase-space}
\end{equation}
which, due to the special form of $\calE$ reduces to
\begin{equation}
    i\dot{u} = 2J^{-1} \hat{H} u, \label{eq:tdse-and-dual-phase-space}
\end{equation}
which is equivalent to Eq.~\eqref{eq:tdse-and-dual}.

\section{Evolution on manifolds}
\label{sec:manifold-evolution}

In this section we describe the bivariational evolution on submanifolds of phase space. We first deal with complex manifolds, where complex differentiation can be used, and then with the more general real manifolds, which will allow us to identify two distinct time-dependent bivariational principles which reduce to the complex case under certain conditions.

\subsection{Complex manifolds}

\begin{figure}
    \centering
    \includegraphics[width=\columnwidth]{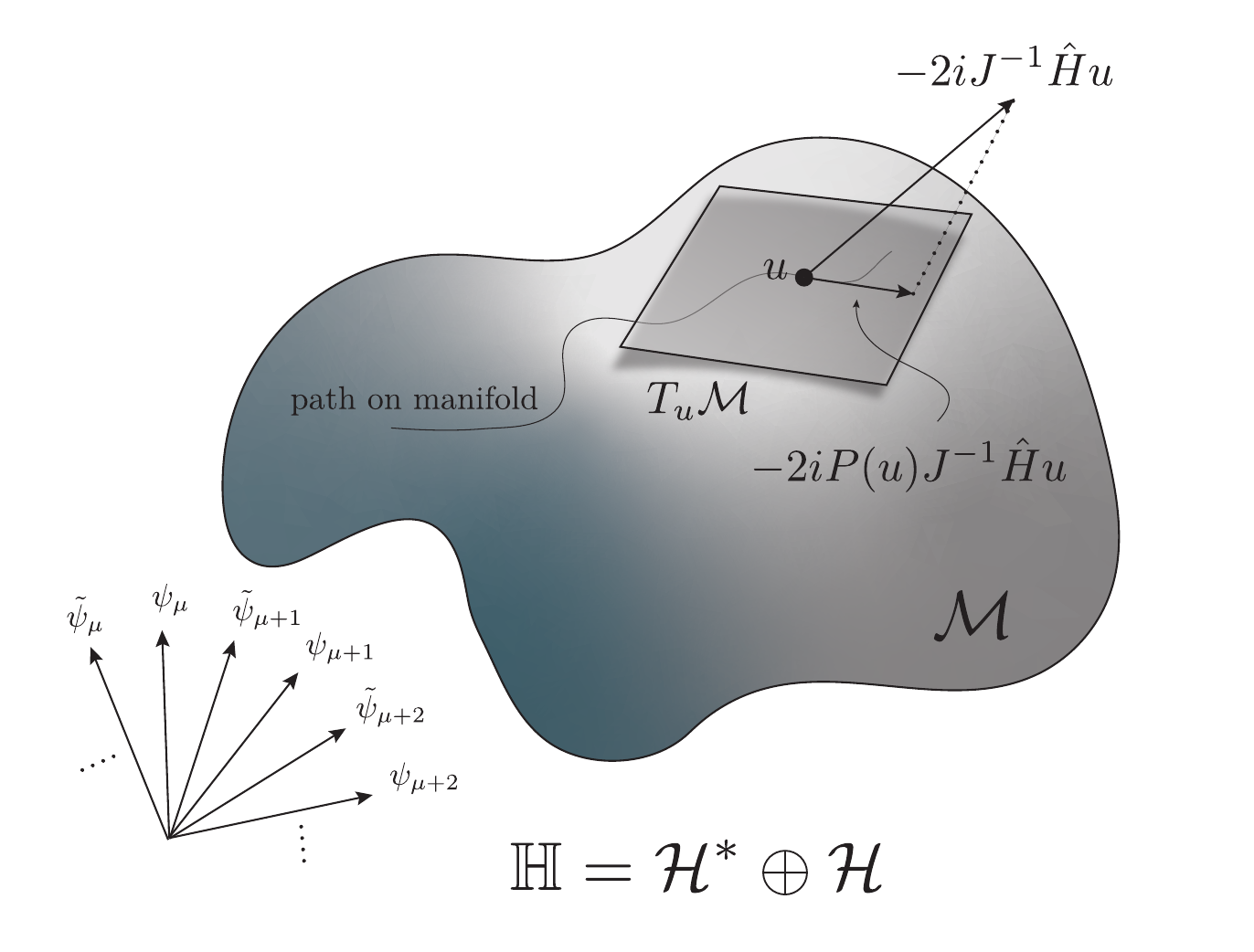}
    \caption{Illustration of infinite dimensional phase space, a submanifold $\calM$, and the symplectic projection that dictates time evolution on the manifold.}
    \label{fig:illustration}
\end{figure}

Approximate time evolution is obtained from the bivariational principle by introducing a smooth submanifold $\calM \subset \HH$ and restricting the principle of stationary action to $\calM$; see Figure~\ref{fig:illustration} for an illustration. We assume for simplicity that $\calM$ is a complex manifold of finite dimension $n$. 

Let $u = (\tilde{\psi},\psi) \in \calM$. Let $\delta u(t) \in T_u\calM$ be an arbitrary variation. We obtain for the variation of the action
\begin{equation}
    \begin{split}
    \delta \calA &= \int i \omega(\delta u, \dot{u}) -  \dualpairing{\delta u,d\calE(u)} \; dt \\
    &= \int i \omega(\delta u, \dot{u}) - \omega(\delta{u}, 2J^{-1} \hat{H} u)\; dt
    \label{eq:variation-action-M}
    \end{split}
\end{equation}
and the corresponding Euler--Lagrange equation
\begin{equation}
    i\dot{u} = P(u) J^{-1}d\calE(u) = 2P(u) J^{-1} \hat{H} u , \label{eq:euler-lagrange-complex}
\end{equation}
where $P(u) : \HH \to T_u\calM$ is the \emph{symplectic projection} at $u$: For any $v\in\HH$, the projection $P(u)v \in T_u\cal M$ is defined by the condition
\begin{equation}
\omega(\delta u, v) = \omega(\delta u, P(u)v) \; \forall \delta u \in T_u\calM.  \label{eq:symp-proj}
\end{equation}
The symplectic projection $P(u)$ is well-defined whenever the restriction $\omega : T_u\calM \times T_u\calM \to \CC$ of the symplectic form is nondegenerate, which is turn is the definition of $\mathcal{M}$ being a symplectic submanifold of $\mathbb{H}$.

\subsection{Real manifolds}\label{Subsection: Real Manifolds}

Suppose now $\calM$ is a \emph{real} manifold of finite dimension $m$. Then, if we consider the functionals
\begin{subequations}
\begin{align}
    \Re \calA &= -\int \frac{1}{2}\Im \omega (u,\dot{u}) + \Re \calE(u) \; dt \label{eq:re-A} \\
        \Im \calA &= \int \frac{1}{2} \Re \omega (u,\dot{u}) - \Im \calE(u) \; dt \label{eq:im-A} ,
\end{align}
\end{subequations}
it no longer holds in general that $\delta\Re\calA=0$ if and only if $\delta\Im\calA=0$. This suggests that we can, in principle, generate \emph{different} approximate time evolutions on $\calM$ from each real-valued action.

The forms $\Re\omega$ and $\Im\omega$ are antisymmetric nondegenerate bilinear forms on a \emph{real} phase space we will denote $\HH_\RR$:
There is a standard way to view a complex linear space $\calV$ as a real linear space $\calV_\RR$, called \emph{the realification of $\calV$}, by restricting the field of multiplicative scalars to $\RR$. The set of vectors is the same. Multiplication $u \mapsto iu$ still yields an element of $\calV_\RR$, but now as a linear operator $\mathbf{i}$ that satisfies $\mathbf{i}^2 = -\operatorname{Id}$. (Such an operator is called a complex structure on the vector space.) In particular,  $u$ and $iu$ are linearly independent in $\calV_\RR$, and the dimension of the space is therefore doubled. Moreover, suppose a nondegenerate bilinear form $a : \calV\times\calV \to \CC$ is given. Both $\Re a$ and $\Im a$ are now nondegenerate bilinear forms on $\calV_\RR$. If $a$ is symmetric/antisymmetric, then $\Re a$ and $\Im a$ are also symmetric/antisymmetric. We conclude that $(\HH_\RR,\Im\omega)$ and $(\HH_\RR,\Re\omega)$ are distinct symplectic linear manifolds.

Select a \emph{real} submanifold $\calM\subset\HH_\RR$, and find, in a similar manner as previously, that $\delta \Re \calA=0$ for all infinitesimal variations $\delta u(t) \in T_u\calM$ if and only if
\begin{subequations}
\begin{equation}
    \dot{u} = -P_{\Im}(u) J^{-1} d\Re \calE,
    \label{eq:euler-lagrange-re-A}
\end{equation}
where $P_{\Im}(u)$ is the symplectic projection operator onto the \emph{real} tangent space $T_u\calM$ obtained from the symplectic form $\Im\omega$. Similarly, $\delta \Im\calA = 0$ for all infinitesimal variations if and only if
\begin{equation}
    \dot{u} = P_{\Re}(u) J^{-1} d\Im \calE,
    \label{eq:euler-lagrange-im-A}
\end{equation}
\end{subequations}
where $P_{\Re}(u)$ is the symplectic projection obtained from the symplectic form $\Re\omega$. The existence of the symplectic projection depends on the invertibility of its matrix in the tangent basis; see Section~\ref{sec:euler-labrange-local}. Equivalently, we must require that $\calM$ is a symplectic submanifold, where the symplectic form is non-degenerate by definition.

The equations of motion~\eqref{eq:euler-lagrange-re-A} and \eqref{eq:euler-lagrange-im-A} are explicitly real equations of motion, and, in the case $\calM = \HH$, equivalent to the canonical equations of motion \eqref{eq:re-ham-1} and \eqref{eq:re-ham-2}.  For general submanifolds, however, the Cauchy--Riemann equations do not apply, and the two Euler--Lagrange equations are not equivalent, i.e., they generate distinct time evolutions on $\calM$. On the other hand, it can happen that $\calM$ is simply a re-expression of a complex manifold using real and imaginary parts of the complex coordinates. In that case, the Cauchy--Riemann equations again apply, and $\Re\calA$ and $\Im\calA$ are equivalent functionals, i.e., Eqs.~\eqref{eq:euler-lagrange-re-A} and \eqref{eq:euler-lagrange-im-A} generate the same time evolution on $\calM$.

\subsection{Manifold normalization}

A word on normalization of bivariational approximation manifolds $\calM$ is in place.
Implicit in Eq.~\eqref{eq:action} is the assumption that $\braket{\tilde{\psi},\psi}=1$, and this normalization is preserved by the time evolution. Indeed, the action $\calA$ can be derived from a more fundamental action principle, see the Appendix, written explicitly on a phase and normalization invariant form using the bivariational density matrix
\begin{equation}
    \rho = \frac{\ket{\psi}\bra{\tilde{\psi}}}{\braket{\tilde{\psi}|\psi}}.
\end{equation}
If the manifold $\calM$ either satisfies $\braket{\tilde{\psi}|\psi}=1$ everywhere, or if $\calM$ contains phase and normalization scalings $(\alpha\tilde{\psi},\beta\psi) \in \calM$ for any fixed $(\tilde{\psi},\psi)\in\calM$, then the two principles of stationary action give the same solutions. In practice, such normalized or scale invariant manifolds are always easy to construct, given a manifold that does not initially satisfy the constraint.

\subsection{Euler--Lagrange equations in local coordinates}
\label{sec:euler-labrange-local}

We express the complex and real Euler--Lagrange equations in terms of local coordinates, beginning with the complex case. Let $\calM\subset\HH$ be a complex submanifold of dimension $n<+\infty$, for simplicity. Let $u \in \calM \subset\HH$ be given in local coordinates by $u = \Phi(z)$, with $z \in \CC^n$, and let $t_\mu = (\tilde{v}_\mu, v_\mu) = \partial_{z^\mu}(\tilde{\psi},\psi) = \partial_{z^\mu} \Phi(z)$ define the coordinate basis vectors. By assumption, this is a linearly independent set. Any tangent vector $\delta u \in T_u\calM$ is now expanded as $\delta u = \sum_\mu t_\mu \delta z^\mu$. The equations of motion are readily obtained by considering Eq.~\eqref{eq:variation-action-M} which leads to the Euler--Lagrange equation
\begin{equation}
    i \Omega_{\mu\nu}\dot{z}^\nu = \frac{\partial\mathcal{E}(z)}{\partial z^\mu} \quad \forall \mu, \label{eq:complex-euler-lagrange-coords}
\end{equation}
where
\begin{equation}
    \frac{\partial\mathcal{E}(z)}{\partial z^\mu} = \braket{\tilde{v}_\mu|H\psi} + \braket{\tilde{\psi}|H v_\mu}, 
\end{equation}
and where
\begin{equation}
    \Omega_{\mu\nu} = \omega(t_\mu, t_\nu) = \braket{\tilde{v}_\mu|v_\nu} - \braket{\tilde{v}_\nu|v_\mu} \label{eq:symplectic form complex euler lagrange}
\end{equation}
is a complex antisymmetric matrix. We used the Einstein summation convention. We note that the antisymmetry implies that (the complex dimension) $\dim(\calM)=n$ must be even. The matrix $\Omega$ is invertible over the manifold if and only if $(\calM,\omega)$ is a symplectic submanifold. 

Inverting $\Omega$ in Eq.~\eqref{eq:complex-euler-lagrange-coords}, multiplying with $t_\nu$ and summing, we obtain the following representation of the symplectic projection operator:
\begin{equation}
    P(u)  = t_\nu (\Omega^{-1})^{\nu\mu}\omega(t_\mu , \; \cdot\; ). \label{eq:symplectic-projection-in-coordinates}
\end{equation}

The calculation of the Euler--Lagrange equations for the real manifold case is very similar. We let $\calM \subset\HH_\RR$ be a real submanifold of dimension $n<+\infty$, and denote as before $(\tilde{v}_\mu, v_\mu) = (\partial_{x^\mu}\tilde{\psi}, \partial_{x^\mu}\psi) \in T_u\calM$ ,  the coordinate basis for tangent space at $u = (\tilde{\psi},\psi)\in\calM$, where now $x = (x^\mu)$ are real-valued local coordinates. We obtain, for $\delta \Re\calA = 0$ and $\delta \Im \calA = 0$, respectively, 
\begin{subequations} 
\label{eq:real-euler-lagrange-coords}
    \begin{align}
         -\Im\Omega_{\mu\nu} \dot{x}^\nu &= \Re \left( \frac{\partial\mathcal{E}(x)}{\partial x^\mu} \right), \quad \text{and} \label{eq:real-euler-lagrange-coords-re-A} \\ 
                 \Re\Omega_{\mu\nu} \dot{x}^\nu &= \Im \left( \frac{\partial\mathcal{E}(x)}{\partial x^\mu} \right) . \label{eq:real-euler-lagrange-coords-im-A}
    \end{align}
\end{subequations}
In both cases, the coefficient matrix is a real antisymmetric matrix. It follows that $\dim(\calM) = n$ must be even both cases. Whenever the matrix inverses exist over the manifold, $(\calM,\Re\omega)$ and $(\calM,\Im\omega)$ are real symplectic manifolds, which must of even (real) dimension. By a similar argument as for Eq.~\eqref{eq:symplectic-projection-in-coordinates}, we derive the following representations for the real symplectic projection operators:

\begin{subequations} 
\label{c}
    \begin{align}
         P_{\Im}(u) &= t_\nu ((\Im \Omega)^{-1})^{\nu\mu}(\Re\omega)(t_\mu , \; \cdot\; ), \label{a} \intertext{and} 
         P_{\Re}(u)  &= t_\nu ((\Re \Omega)^{-1})^{\nu\mu}(\Im\omega)(t_\mu , \; \cdot\; ).  \label{v}
    \end{align}
\end{subequations}

\subsection{Interpretation}

Taking the real part of $\calA$ to define an action functional is not a novel idea, and dates back at least to Pedersen and Koch~\cite{pedersen_time-dependent_1998}. The time-dependent optimized-orbital CC method of Sato and coworkers~\cite{satoCommunicationTimedependentOptimized2018} (see also Section~\ref{sec:current-methods}) is defined in terms of taking the real part of a non-complex differentiable action.
Indeed, for bivariational methods such as the coupled-cluster method, it is a well-known problem, or feature, that observables may attain non-zero imaginary values. It is customary to simply insist on taking the real part of the computed observable, and discard the (hopefully) small imaginary parts. This would be equivalent to using the action $\Re\calA$ and using the Hellmann--Feynman interpretation of expectation values~\cite{pedersen_time-dependent_1998}. Indeed, for $\Re \calA$, the real part $\Re\calE$ of the energy expectation value is now the generator for the dynamics, and for a generic observable $O$, the expectation value functional reads $\Re \calE_O(u) = \Re (\dualpairing{u,\hat{O}u}/\dualpairing{u,\hat{I}u})$, where $\hat{I}$ is the symmetrization of the identity operator.

On the other hand, we have also found that taking the \emph{imaginary} part $\Im\calA$ yields a distinct approximation when $\calM$ is real. Then, $\Im\calE$ is the generator for dynamics and hence conserved. 

Using the imaginary part of the energy as generator for dynamics may seem odd. However, consider the following: For $\Re\calA$, the real part  $\Re\calE$ is conserved in time, while we have no conservation law for $\Im\calE$. If the manifold $\calM$ is accurate, we can expect $\Im\calE$ to remain small. On the other hand, for $\Im\calA$, the imaginary part $\Im\calE$ is conserved -- and if $\calM$ is accurate, it will be small -- but it is now $\Re\calE$ that fluctuates, but should hopefully be \emph{almost} conserved. In this sense, the two principles are complementary, and reflect the ubiquitous compromise in bivariational theory, which is ``non-Hermitian'' in nature.

\subsection{Poisson brackets and conservation laws}

\subsection{Poisson bracket}

Let $\mathcal{F}(u)$ and $\mathcal{G}(u)$ be smooth scalar-valued functions of $u \in \mathcal{M}$, for the moment assumed to be a complex submanifold of $\mathbb{H}$. We define a Poisson bracket in local coordinates,
\begin{equation}
    \{\mathcal{F},\mathcal{G}\} := -i (\nabla_z \mathcal{F})^T \Omega^{-1} \nabla_z \mathcal{G}, \label{eq:poisson in local coordinates}
\end{equation}
which is again a smooth scalar valued function over $\mathcal{M}$.  Due to the antisymmetry of $\omega$, 
\begin{equation}
    \{\mathcal{F},\mathcal{G}\} = -\{\mathcal{G},\mathcal{F}\},
\end{equation}
and it is also readily shown\cite{lubich_quantum_2008} that the bracket satisfies the Jacobi identity
\begin{equation}
    \{\mathcal{F},\{\mathcal{G},\mathcal{H}\}\} + 
    \{\mathcal{G},\{\mathcal{H},\mathcal{F}\}\} + 
    \{\mathcal{H},\{\mathcal{F},\mathcal{G}\}\} = 0.
\end{equation}
In the case where $\mathcal{M} = \mathbb{H}$, the Poisson bracket takes the form 
\begin{equation}
    \{\mathcal{F},\mathcal{G}\} = i \dualpairing{J^{-1} d\mathcal{F}, d\mathcal{G}}, \label{eq:poisson-bracket-exact}
\end{equation}
which is, naturally, coordinate-free. By using the representation~\eqref{eq:symplectic-projection-in-coordinates} we obtain a coordinate-free formula for the Poisson bracket~\eqref{eq:poisson in local coordinates} on the complex manifold $\mathcal{M}$
\begin{equation}
    \{\mathcal{F},\mathcal{G}\} = i \dualpairing{P(u)J^{-1}d\mathcal{F},d\mathcal{G}}. \label{eq:poisson on manifold without coordinates}
\end{equation}
The Poisson bracket $\{\cdot, \calE\}$ generates time evolution. By the chain rule, it is seen that
\begin{equation}
    \frac{d \mathcal{F}}{dt} = \{\mathcal{F}, \mathcal{E}\}. \label{Equation: dF/dt}
\end{equation}
In particular, the coordinates themselves are smooth over $\mathcal{M}$, which gives 
\begin{equation}
    \frac{d z^\mu}{dt} = \{ z^\mu, \mathcal{E}\}.
\end{equation}

\subsection{Ehrenfest's Theorem}

Recall that for an operator  $A$  on $\mathbb{H}$, we have defined its symmetrization $\hat{A}$ on $\mathbb{H}\rightarrow \mathbb{H}^*$ as $\hat A(\tilde\psi,\psi) = \tfrac{1}{2} (A\psi,A^t\tilde\psi)$. We invite the reader to verify the following useful formulas: For all bounded operators $A,B : \mathbb H \to \mathbb H^*$, and for all $u,v\in\mathbb H$, $w\in \mathbb H^*$
\begin{subequations} 
\label{Equation: Magic Commutator Formula}
\begin{align}
\llangle u, \hat A v\rrangle &= \llangle v,\hat Au\rrangle,\label{Equation: Magic Commutator Formula a}\\
\llangle u,w\rrangle & = -\llangle J^{-1} w, Ju\rrangle \label{Equation: Magic Commutator Formula e}\\ 
\intertext{and}
\llangle J^{-1}\hat Au,\hat Bu\rrangle &= -\frac{1}{4}\llangle u,\widehat{[A,B]}u\rrangle.\label{Equation: Magic Commutator Formula g}
\end{align}\end{subequations} 
We define the expectation value of $A$ with respect to $(\tilde\psi,\psi)\equiv u\in\mathbb H$ by 
\[\llangle A\rrangle_u := \dualpairing{u,\hat A u} = \langle \tilde \psi| A \psi\rangle. \]
Let $u(t)$ be a solution to Eq.~\eqref{eq:tdse-and-dual-phase-space}. Then, using Eq.~\eqref{Equation: Magic Commutator Formula g} and the Poisson bracket~\eqref{eq:poisson-bracket-exact} with $d\llangle  A\rrangle_u = 2\hat Au$, one obtains the following Poisson bracket:
\begin{equation}
    \{\dualpairing{A}_u, \dualpairing{B}_u\} = -i \dualpairing{[A,B]}_u.
\end{equation}
In particular, one obtains Ehrenfest's Theorem on bivariational form,
\begin{equation}
    \frac{d}{dt}\llangle  A \rrangle_u = -i\llangle {[A,H]}\rrangle_u.
    \label{Equation: Bivariational Ehrenfest}
\end{equation}

\subsection{Conservation laws on complex manifolds}

We next present the generalization of the  Bivariational Ehrenfest theorem to manifolds. Let $\mathcal M \subset \mathbb H $ be a complex manifold. We say that $\hat A$ \emph{preserves} $\mathcal M$ if 

\[\forall u\in \mathcal M: J^{-1}\hat A  u\in T_u \mathcal M.\]
Let $u(t)=(\tilde\psi(t),\psi(t))\in \mathcal{M}$ be a solution to the bivariational principle on $\mathcal M$, and hence a solution to the $\mathcal M$-projected Euler--Lagrange equation, Eq.~\eqref{eq:euler-lagrange-complex}. 

Using Eq.~\eqref{Equation: dF/dt}, we see that 
\[\frac{d}{dt}\llangle \hat A\rrangle_u = \{\llangle A\rrangle_u,\calE\}.\]
Applying Eq.~\eqref{Equation: Magic Commutator Formula e} to the Poisson bracket formula in Eq.~\eqref{eq:poisson on manifold without coordinates}, and inserting $d\llangle \hat A\rrangle_u = 2\hat Au$ and $d\mathcal E = 2\hat Hu,$ we get
\[\frac{d}{dt}\llangle \hat A\rrangle_u = -4i\llangle J^{-1}\hat Au,JP(u)J^{-1} \hat Hu\rrangle \]
If $\hat A$ preserves $\calM$, then we can use the definition of $P(u)$ in Eq.~\eqref{eq:symp-proj} together with Eq.~\eqref{Equation: Magic Commutator Formula g} to obtain
\[\frac{d}{dt}\llangle \hat A\rrangle_u = -i\llangle [A,H]\rrangle_u \]
In general, $\{\llangle A\rrangle_u,\llangle B\rrangle_u\}=-i\llangle [A,B]\rrangle_u$
if $\hat A$ or $\hat B$ preserves $\calM$. If $\hat A$ does not preserve $\calM$, it is possible to express the deviation from the Ehrenfest theorem in terms of the distance between $J^{-1}\hat Au$ and $T_u\calM$. One can thus obtain bounds for change in expectation value of observables. Indeed, we have in general
\[\{\llangle A\rrangle_u,\llangle B\rrangle_u\} = -i\llangle [A,B]\rrangle_u\]\[ -4i \llangle u,\hat B Q(u)J^{-1}\hat Au\rrangle, \]
where $Q(u)=1-P(u)$. Thus, for a preserved variable $\llangle A\rrangle $ evolving on a complex manifold, we obtain the bound of the error 
\[\int \bigg|\frac{d}{dt}\llangle A\rrangle\bigg| dt \leq 4 \int| \llangle u,\hat A Q(u) J^{-1}\hat H u\rrangle|dt\]
Such error bounds can be expressed in terms of the curvature of the manifold $\calM,$\cite{lubich_quantum_2008} but we do not consider this further here.

\section{Relation to univariational theory}

\subsection{TDVP vs.~McLachlan variational principle}

The branching of the complex TD-BIVP into two distinct real bivariational principles is similar to the relationship between the time-dependent variational principle (TDVP) and the McLachlan variational principle for \emph{univariational} approximations of the Schrödinger equation~\cite{kramer_geometry_1981,broeckhove_equivalence_1988,lubich_quantum_2008,hackl_geometry_2020,yuan_theory_2019}. The TDVP recasts the TDSE as a principle of stationary action via the functional
\begin{equation}
    \mathcal{S} =\int \braket{\psi(t), \dot{\psi}(t) + i H\psi(t)} \, dt\label{eq:tdvp-action}
\end{equation} 
where $\braket{\cdot,\cdot}$ is the ordinary inner product.
Using the fact that $H$ is Hermitian it is straightforward to show that $\delta \mathcal{S}=0$ if and only if $i\dot{\psi}=H\psi$. Suppose now $\mathcal{M}\subset\mathcal{H}$ is a \emph{real} submanifold. The stationary conditon of $\mathcal{S}$ now becomes after using integration by parts
\begin{equation}
    \Im \langle\delta\psi,\dot{\psi}+iH\psi\rangle = 0 \label{eq:tdvp}
\end{equation}
for all $\delta \psi \in T_\psi \mathcal{M}$. 

If we define the symplectic form $\tilde{\omega}(\psi,\phi) = 2i\Im \langle \psi,\phi\rangle $ on $\mathcal{H}$, then, Eq.~\eqref{eq:tdvp} is equivalent to 
\[\tilde\omega(\delta\psi,\dot\psi+iH\psi)=0\;\forall \psi\in T_\psi\mathcal M. \]
We may rephrase in terms of the symplectic projection $P_{\tilde\omega}(\psi):\mathcal H\rightarrow T_\psi\mathcal M$, such that Eq.~\eqref{eq:tdvp} is equivalent to
\begin{equation}
\dot{\psi}=P_{\tilde{\omega}}(\psi)(-iH\psi).
\label{eq:tdvp-projection}
\end{equation}

The McLachlan variational principle\cite{mclachlan_variational_1964,lubich_quantum_2008} states that  $\dot{\psi}=\Theta$, where $\Theta$ is defined by
\begin{equation*}
\Theta = \argmin_{\Theta\in T_\psi \mathcal{M}}\|iH\psi+\Theta\|^2.
\end{equation*}
Differentiating $\|iH\psi+\Theta\|^2$ with respect to $\Theta$, equating to zero and inserting $\dot{\psi}=\Theta$ gives
\begin{equation}
    \Re\langle \delta \psi,\dot{\psi} + iH\psi\rangle =0, 
    \label{Equation: MacLachlan 1}
\end{equation}
for all $\delta \psi\in T_\psi\mathcal{M}$. The form $\Re\braket{\psi,\phi}$ is the inner product on the realification $\mathcal{H}_\RR$, so that Eq.~\eqref{Equation: MacLachlan 1} is equivalent to
\begin{equation}
    \dot{\psi} = P_\bot(\psi)(-iH\psi), \label{eq: McLachlan projector}
\end{equation}
with $P_\bot(\psi) : \calH_\RR \to T_\psi\calM$ being the orthogonal projection.

Suppose now $\mathcal{M}$ has the property that $\delta\psi\in T_\psi\calM\iff i\delta\psi \in T_\psi\calM$. This happens if $\calM$ is actually a complex submanifold of $\calH$. Then Eq.~\eqref{Equation: MacLachlan 1} and Eq.~\eqref{eq:tdvp} are equivalent, and the symplectic and orthogonal projections coincide, $P_{\tilde{\omega}}(\psi) = P_\bot(\psi)$.

The TDVP and the McLachlan principles are distinct for real manifolds, but equivalent for complex manifolds, where the symplectic projection and orthogonal projections coincide. We contrast this with the bivariational case, where the projections $P_{\Re}(u)$ and $P_{\Im}(u)$ are distinct for real manifolds, and $P_{\Re}(u) = P_{\Im}(u)=P(u)$ for complex manifolds. Notably, in all cases we deal with the symplectic projection. 

\subsection{TD-BIVP contains the TDVP, but not the McLachlan principle}

Consider the approximate univariate dynamics on a real or complex submanifold $\calM\subset\calH$ generated by the TDVP (Eq.~\eqref{eq:tdvp-projection}). We denote by $\Phi_t : \calM \to \calM$ the TDVP flow, such that given an initial condition $\psi(0) \in \calM$, we have $\psi(t) = \Phi_t(\psi(0))$. 

Let $j : \calH \to \mathbb{H}$ be the map $j(\psi) = (\bar{\psi},\psi)$, where $\bar{\psi}$ is the injection of $\calH$ into $\calH^*$. (It is not the complex conjugate, which does not exist for general complex Hilbert spaces, but rather a ``renaming'' of $\psi$ so that it lies in $\calH^*$.\cite{budinichSpinorialChessboard1988}) A straightforward calculation shows that
\begin{equation}
    \omega(j(\psi),j(\phi)) = \tilde{\omega}(\psi,\phi),
\end{equation}
thereby relating the symplectic structure on $\calH$ to that of $\mathbb{H}$. Moreover, let $\mathcal{N} =  j[\mathcal{M}]$. We have $T_{j(\psi)}\mathcal{N} = j[T_\psi\mathcal{M}]$ for any $\psi \in \calM$. The manifold $\mathcal{N}$ is complex if and only if $\mathcal{M}$ is complex. The symplectic form $\omega$ on $\mathbb{H}$ is purely imaginary on $j[\mathcal{M}]$, so we select the real symplectic form $\Im\omega$ and thus the  
real part of the bivariational action, Eq.~\eqref{eq:re-A}, and generate a flow $\Psi_t : \mathcal{N} \to \mathcal{N}$ such that $u(t) = \Psi_t(u(0))$.

It is now a straightforward calculation to show that the following diagram commutes:
\begin{center}
\begin{tikzcd}
\psi(0) \arrow[d, "j"'] \arrow[r, "\Phi_t"]   & \psi(t) \arrow[d, "j"] \\
u(0) \arrow[r, "\Psi_t"'] & u(t) 
\end{tikzcd}
\end{center}
In other words, the TDVP is contained in the TD-BIVP on the form $\Re \delta\calA = 0$, when promoting a real manifold $\mathcal{M}$ to $j[\mathcal{M}]$ as bivariational manifold. If $\mathcal{M}$ is complex, we may use any bivariational action, real or complex.

We now turn to the McLachlan principle, which uses an orthogonal projection. For a complex submanifold $\mathcal{M}\subset \calH$, the McLachlan principle and the TDVP coincide. If, however, $\mathcal{M}$ is a real manifold, $P_\bot(\psi)$ is not a symplectic projection, indicating that the McLachlan principle is not contained in the BIVP in this case. Indeed, we instead find that it is a special case of the following bivariational version of the McLachlan principle:
\begin{equation}
    \Delta = \dualpairing{\delta u, \hat{I}(\dot{u} + 2 i {J}^{-1} \hat{H} u)} = 0, \quad \forall \delta u \in T_u \mathcal{N}, \label{eq:bivp-mclachlan}
\end{equation}
valid also in the \emph{exact} case, and hence we can elevate it to a general principle valid also for general manifolds $\mathcal{N}\subset\mathbb{H}^*\oplus\mathbb{H}$.

The principle~\eqref{eq:bivp-mclachlan} does not seem to be derivable from a stationary action. The reason is that the first term $\dualpairing{\delta u, \hat{I}\dot{u}}$ is a symmetric bilinear form, and the functional $\int \dualpairing{u,\hat{I}\dot{u}} dt$ is therefore a total time derivative.

To derive Eq.~\eqref{eq:bivp-mclachlan}, we note that
\begin{equation}
\begin{split}
 \Delta &\equiv \Re \braket{\delta\psi,\dot{\psi} + i H\psi} = \Im \braket{\delta\psi,i\dot\psi - H\psi} \\&= \frac{1}{2i} \omega(j(\delta\psi),j(i\dot{\psi} - H\psi)).
 \end{split}
\end{equation}
For a point $u = j(\psi) \in \mathcal{N}$, we now have
the relations $\dot{\psi} = j^{-1}\dot{u}$ and $jH \psi = \hat{I}^{-1}\hat{H}u$, verified by direct computation. Moreover, $j i j^{-1} = \tfrac{1}{2}i \hat{I}^{-1} J$, also verified by direct computation. This results in
$$
 \Delta = \frac{1}{2} \omega(\delta u, \hat{I}^{-1}(J\dot{u} + 2i \hat{H}u)).
$$
At this point, we insert the definition of the symplectic form, with some easy manipulations, to obtain Eq.~\eqref{eq:bivp-mclachlan}.

\section{Unifying view of current time-dependent bivariational approaches}
\label{sec:current-methods}

The abstract formalism in this article is applicable to many methods for real-time propagation encountered in the literature. In this section, we give a brief overview of what we consider to be some important examples in the language of the present work, all being varieties of coupled-cluster (CC) theory: Time-dependent traditional CC theory (TDCC), orbital-adaptive time-dependent CC (OATDCC) theory, orthogonal orbital-optimized time-dependent CC (TD-OCC) theory, and time-dependent equation-of-motion CC (TD-EOM-CC) theory. 

Familiarity with CC theory is assumed, and this section will only serve as an overview. We refer to the original publications for full details and complete specifications. For a recent review of real-time propagation with CC theory, Ref.~\onlinecite{sverdrup_ofstad_time-dependent_2023}.

\subsection{The traditional CC ansatz}

The traditional CC method is the most popular wavefunction-based method for electronic-structure theory, with the CCSD(T) model often termed ``the gold standard of quantum chemistry'' due to its balance of cost and accuracy\cite{bartlettCoupledclusterTheoryQuantum2007}.
Similarly, the CC method offers an attractive approach for approximating the solution to the vibrational Schrödinger equation\cite{christiansenVibrationalCoupledCluster2004}.  Although the physical nature of the degrees of freedom and of the Hamiltonian is very different in the electronic and vibrational cases, both cases benefit from fast convergence of the CC hierarchy, polynomial-scaling cost, and size extensivity.

The traditional CC method is usually formulated in finite-dimensional subspace of $\calH$, defined in terms of a finite orthonormal set of single-particle functions (a ``basis set''). However, we here take a broader picture, and merely assume a \emph{biorthogonal} set of  single-particle functions partitioned into occupied and unoccupied subsets, $\mathcal{B} = \mathcal{B}_\text{occ}\cup\mathcal{B}_\text{unocc}$, and $\tilde{\mathcal{B}} = \tilde{\mathcal{B}}_\text{occ}\cup\tilde{\mathcal{B}}_\text{unocc}$.. This induces a biorthogonal many-particle basis set $\{\phi_\mu\}\subset \calH$ and $\{\tilde{\phi}_\mu\}\subset \calH^*$ with $(\tilde{\phi},\phi) = (\tilde{\phi}_0,\phi_0)$ being formal excitation references for bras and kets, respectively. The full configuration-interaction wavefunction and its dual are written $\psi = C \phi$ and $\tilde{\psi} = \tilde{C}^t \tilde{\phi}$, with $C = \sum_\mu c_\mu X_\mu\in \calC$  and $\tilde{C} = \sum_\mu \tilde{c}_\mu \tilde{X}_\mu\in\tilde{\calC}$ being cluster operators (the summations run over $\mu \geq 0)$. Here, $X_\mu \phi = \phi_\mu$ is an elementary excitation, and similarly $\tilde{X}_\mu^t \tilde{\phi} = \tilde{\phi}_\mu$ is an elementary dual excitation, or de-excitation. 
Note that $X_0 \phi = \phi$ and $\tilde{X}_0^t \tilde{\phi} = \tilde{\phi}$, i.e. $X_0$ and $\tilde{X}_0$ are simply identity operators or null excitations. The notation $A^t$ is defined via the dual pairing $\braket{A^t\phi|\psi} \equiv \braket{\phi|A\psi}$. Equivalently, using bra notation, $\bra{A^t\tilde{\phi}} = \bra{\tilde{\phi}}A$.

The spaces $\tilde{\calC}$ and $\calC$ are nilpotent abelian algebras that depend on the subdivisions $\mathcal{B} = \mathcal{B}_\text{occ}\cup\mathcal{B}_\text{unocc}$ and $\tilde{\mathcal{B}} = \tilde{\mathcal{B}}_\text{occ}\cup\tilde{\mathcal{B}}_\text{unocc}$ of the single-particle basis set into occupied and unoccupied subsets. The spaces $\tilde{\mathcal{C}}$ and $\mathcal{C}$ are duals to each other, with the dual pairing being given by $\braket{\tilde{\phi}|\tilde{C}C\phi} = \sum_{\mu} \tilde{c}_\mu c_\mu$.

In traditional CC theory,  the phase-space point $(\tilde{\psi},\psi)$ is parameterized in terms of a pair of cluster operators $(\Lambda,T) \in\tilde{\calC} \oplus \calC$ as
\begin{equation}
    \psi = e^T\phi, \quad \tilde{\psi} = e^{-T^t} \Lambda^{\!t} \tilde{\phi}
    \label{eq:traditional-cc},
\end{equation}
where $T = \sum_\mu \tau_\mu X_\mu$ and $\Lambda = \sum_\mu \lambda_\mu \tilde{X}_\mu$. Again, $\mu = 0$ is included in the summations. $\tau_0$ plays the role of a phase/norm factor, while $\lambda_0$ determines the intermediate normalization in the sense $\braket{\tilde{\psi}|\psi} = \lambda_0$.
Equation~\eqref{eq:traditional-cc} defines a map $\Phi(\Lambda,T) = (\tilde{\psi},\psi)$ being a global coordinate chart for a smooth complex submanifold $\calM_{\mathrm{CC}} \subset \HH$.  In fact, this submanifold is covers \emph{almost} all possible points in $\HH$. The additional conditions are $\braket{\tilde{\psi}|\psi} \neq 0$ and $\braket{\tilde{\phi}|\psi} \neq 0$.

The energy functional in these coordinates is
\begin{equation}
    \calE(\Lambda,T) = \braket{\tilde{\phi}| \Lambda e^{-T} H e^{T} \phi},
\end{equation}
the conventional CC Lagrangian (which is a Lagrangian in the sense of constrained optimization, and must not to be confused with the Lagrangian density encountered in the bivariational principle), and the action functional reads
\begin{equation}
    \calA_\text{CC}= \int i \lambda\cdot\dot{\tau} - \calE(\Lambda,T)\; dt.
    \label{eq:CC action}
\end{equation}
(The integrand is the Lagrangian density in our language.) In particular, the functional form is preserved compared to Eq.~\eqref{eq:action}. This means, that the coordinate transformation $(\tilde{\psi},\psi) = \Phi(\Lambda,T)$ is a canonical transformation in the sense of classical mechanics, and it follows that Hamilton's equations of motion (both the complex and real forms) are preserved as well. 

The full CC case is not practical, and conventional truncation schemes $\mathcal{T}$ of the cluster operators imply an approximate submanifold $\calM_{\mathrm{CC}}(\mathcal{T}) \subset \calM_{\mathrm{CC}}$.  For electronic-structure theory, $\mathcal{T}$ is typically the $\mathrm{CCSD} \cdots K$ scheme, where all excitations of up to $K$ electrons are included.
In the vibrational case, the analogous approach is usually denoted VCC$[K]$ and includes up to $K$-mode excitations.
Since the coordinates are canonical, the induced symplectic form on $\calM_{\mathrm{CC}}(\mathcal{T})$ is trivially non-degenerate, and the submanifold is always symplectic.

Since the coordinates $(\Lambda,T)$ are canonical, the Poisson bracket takes on the simple form
\begin{equation}
    \begin{split}
    \{\mathcal{F},\mathcal{G}\} &= i\dualpairing{J^{-1}d\mathcal{F},d\mathcal{G}} \\&= i\left(\left\langle{\frac{\partial{\mathcal{G}}}{\partial \tau}\Big| \frac{\partial{\mathcal{F}}}{\partial \lambda}}\right\rangle - \left\langle{\frac{\partial{\mathcal{F}}}{\partial \tau}\Big| \frac{\partial{\mathcal{G}}}{\partial \lambda}}\right\rangle \right).
    \end{split}
\end{equation}
For expectation values, $\mathcal{F} = \braket{\tilde{\psi}|F\psi} = \braket{\tilde{\phi}|\Lambda e^{-T} F e^T|\phi}$, we have $\partial\mathcal{F}/\partial {\tau^\mu} = \braket{\tilde{\phi}|\Lambda [\bar{F},X_\mu]|\phi}$, where $\bar{F} = e^{-T}He^T$, and $\partial\mathcal{F}/\partial{\lambda^\mu} = \braket{\tilde{\phi}_\mu|\bar{F}\phi}$. Evaluation of the Poisson bracket yields
\begin{multline}
    \{\mathcal{F},\mathcal{G}\} = -i\braket{\tilde{\psi}|[F,G]\psi} \\+ i\sum_\mu (\braket{\tilde{\phi}|\Lambda X_\mu \bar{G}|\phi}\braket{\tilde{\phi}_\mu|\bar{F}|\phi} \\- \braket{\tilde{\phi}|\Lambda X_\mu \bar{F}|\phi}\braket{\tilde{\phi}_\mu|\bar{G}|\phi})  
\end{multline}
In particular, if we set $\mathcal{F}=\calE$, we obtain that $\mathcal{G}$ is exactly conserved under dynamics only if the last two terms vanish.

\subsection{The OACC ansatz}
The CC ansatz described above is defined in terms of a \textit{static}
single-particle basis. In particular, the references $(\tilde{\phi}, \phi)$ are static, which is a serious limitation in terms of describing, say, large oscillations in the wavefunction, or motion far away from the ground state. The traditional CC ansatz
works well when the amplitudes are sufficiently small, i.e., when the reference describes a large part of the wave function.
Conversely, if the wave function moves too far from the reference, the amplitudes grow and the ansatz tends to break down.
The paradigmatic example of such a situation is ionization or dissociation, i.e. the removal of one or more particles from the system (typically by a laser pulse).
However, much less violent phenomena can also initiate the breakdown of the CC ansatz, as exemplified in vibrational CC theory by the internal vibrational energy redistribution (IVR) in water\cite{madsen_time-dependent_2020}.

This problem can be alleviated by introducing an \textit{adaptive} single-particle basis: Both the occupied and unoccupied single-particle bra and ket basis functions are time dependent. This in turn defines a pair of adaptive references \emph{and} excited determinants that move with the wave function, so to speak. In the present exposition, we assume that the span of the single-particle bases $\mathcal{B}$ and $\tilde{\mathcal{B}}$ are fixed. Equivalently, there is no \emph{secondary space}, and the CC parameterizaton is allowed to correlate all available single-particle functions. This means that the OACC ansatz becomes formally equivalent to non-orthogonal orbital-optimized CC theory (NOCC)\cite{bondopedersenGaugeInvariantCoupled2001}.

Consider therefore a pair of uncorrelated states  $(\tilde{\phi},\phi) \in \calH^* \oplus\calH$ satisfying $\braket{\tilde{\phi}|\phi}=1$. The set of such binormalized reference determinants is a smooth submanifold $\mathcal{U}\subset \mathbb{H}$. Every element $(\tilde{\phi},\phi)\in\mathcal{U}$ defines a unique CC manifold $\mathcal{M}_\text{CC}(\tilde{\phi},\phi)$.
The complex orbital-adaptive CC manifold is defined by
\begin{equation}
    \mathcal{M}_{\mathrm{OACC}}(\mathcal{T}) = \bigcup_{(\tilde{\phi},\phi) \in \mathcal{U}} \mathcal{M}_{\mathrm{CC}} (\tilde{\phi},\phi;\mathcal{T}).
\end{equation}
It has been found that singles excitations must be removed from $\mathcal{T}$ in order to produce a well-defined manifold\cite{kvaal_ab_2012}.

The manifold $\mathcal{M}_\text{OACC}(\mathcal{T})$ is a complex manifold due to the independence of the bra and ket single-particle functions.

The orbital-adaptive time-dependent coupled-cluster (OATDCC) ansatz was first introduced by Kvaal\cite{kvaal_ab_2012} for describing electron dynamics. An analogous ansatz for the vibrational problem was proposed by Madsen et al.\cite{madsen_general_2020} under the name time-dependent modal vibrational coupled-cluster (TDMVCC).
The formulation by Kvaal is done in an abstract infinite-dimensional setting, but,
for simplicity, we consider here a finite basis and use the exponential parameterization of the manifold $\mathcal{U}$ of reference determinants. To that end, given  an arbitrary $(\tilde{\phi},\phi)\in \mathcal{U}$, any \emph{other} $(\tilde{\phi}',\phi')\in\mathcal{U}$ can be written written as
\begin{align}
    \phi' = e^{\kappa} \phi, \quad \tilde{\phi}' = e^{-\kappa^t} \tilde{\phi}.
\end{align}
Here, $\kappa$ is a generic (neither Hermitian nor anti-Hermitian) one-particle operator,
\begin{align}
    \kappa = \sum_{pq} \kappa_{pq} \tilde{c}_p^\dagger c_q
\end{align}
We have introduced the creation and annihilation operators associated with the biorthogonal single-particle basis of the traditional CC method.  In fact, $\exp(\kappa)$ and $\exp(-\kappa^t)$ are change of single-particle basis operators. Correspondingly, for any $(\tilde{\psi},\psi) \in \mathcal{M}_{\text{CC}}(\tilde{\phi},\phi;\mathcal{T})$, we can apply the basis change operators and obtain \emph{every} element 
\begin{equation}
    (\tilde{\psi}',\psi') = (e^{-\kappa^t}\tilde{\psi},e^\kappa\psi) \in \mathcal{M}_\text{CC}(\tilde{\phi}',\phi';\mathcal{T}).
\end{equation}

However, the parameterization contains redundancies that must be eliminated or fixed by a suitable gauge condition. The source of the redundancy is the invariance of the CC wavefunctions under mixings of occupied single-particle functions and unoccupied single-particle functions separately. Thus, one allowed gauge condition (at $\kappa=0$) is to only keep all elements in $\kappa$ that mix occupied single-particle functions [those that comprise $(\tilde{\phi},\phi)$] with unoccupied single-particle functions, and set all other elements to zero. The mathematical structure is that of a principal bundle\cite{lubich_quantum_2008}.

It is instructive to consider
the Lagrangian, i.e. the integrand of the action, Eq.~\eqref{eq:action}:
\begin{equation}
    \begin{split}
    \mathcal{L}_\mathrm{OACC}
    &= 
    i\braket{\tilde{\psi}'|\dot{\psi}'} -  \braket{\tilde{\psi}'|H\psi'} \\
    &= i\braket{\tilde{\psi}|\dot{\psi}} -  \braket{\tilde{\psi}|(\bar{H} - G)\psi}.
\end{split}
\end{equation}

Here, $\bar{H} = e^{-\kappa} H e^{\kappa}$ and $G = i e^{-\kappa} (de^{\kappa}/dt) = -i (de^{-\kappa}/dt) e^{\kappa}  $. $\mathcal{L}_\mathrm{OACC}$ is in fact identical to the integrand of the action in traditional CC theory, cf.~Eq.~\eqref{eq:CC action}, provided we substitute $H \leftarrow \bar{H} - G$, so the AOCC amplitude equations have the same form as the CC amplitude equations.\cite{kvaal_ab_2012,madsen_general_2020}
Stationarity of $\mathcal{A}_\mathrm{OACC}$ leads to a set of linear equations for $G$, which in turn determines $\dot{\kappa}$. At the point $\kappa = 0$, the relation is particularly simple, namely $G = i \dot{\kappa}$ (the $\kappa \neq 0$ case has been treated in detail in Ref.~\citenum{hojlundGeneralExponentialBasis2023a}). We remark that Refs.~\citenum{kvaal_ab_2012} and \citenum{madsen_general_2020} did not use the exponential parameterization for the single-particle basis, but the linear equations are the same.

\subsection{Orthogonal orbital-optimized CC}

The orthonormal orbital-optimized CC (OCC) ansatz is identical to the OACC ansatz except that the single-particle basis is restricted to being orthonormal. This induces an orthonormal many-particle basis and, in particular, $\tilde{\phi} = \phi^\dag$. We thus obtain a \textit{real} submanifold,
\begin{equation}
    \mathcal{M}_{\mathrm{OCC}} = \bigcup_{(\phi^\dag,\phi)\in\mathcal{U}} \mathcal{M}_{\mathrm{CC}} (\phi^\dag,\phi).
\end{equation}
Whereas $\mathcal{M}_\text{OACC}$ was a complex manifold, the OCC manifold is a \emph{real} manifold, which can be seen by realizing that the ``coordinates'' being single-particle basis functions appear both as complex conjugates and as they are. Consequently, one must turn to one of the real action principles $\delta \Re \mathcal{A}_\text{OCC} = 0$ or
$\delta \Im \mathcal{A}_\text{OCC} = 0$, which lead to distinct time evolutions. Sato et al.~\cite{satoCommunicationTimedependentOptimized2018} used $\Re \mathcal{A}$, which amounts to letting $\Re \calE$ generate the time evolution. As the physical energy is a real quantity, this seems to be the natural choice.

The CC amplitudes $\lambda$ and $\tau$ appears in a complex differentiable manner in the OCC ansatz, so $\Re \mathcal{A}_\text{OCC}$ leads to the same amplitude equations as $\mathcal{A}_\text{OCC}$. The OCC and OACC amplitude equations are thus identical, and both are essentially identical to the traditional CC amplitude equations. The linear equations that determine the OCC basis set evolution are, however, symmetrized in OCC compared to the OACC equations. As an example, the density matrices that appear in the OACC working equations are Hermitianized in the OCC working equations (cf. Refs.~\citenum{satoCommunicationTimedependentOptimized2018} and \citenum{kvaal_ab_2012}).

One may consider appealing to the action principle $\delta \Im\mathcal{A}_\text{OCC}$. This will again lead to the same amplitude equations as before, while the basis set equations will contain anti-Hermitianized density matrices etc. While it seems like an unconventional choice, it may be worth investigating in future research.

\subsection{Time-dependent EOM-CC}

Equation-of-motion CC theory is perhaps the simplest example of a bivariational method that is not explicitly Hermitian. In EOM-CC, a ground-state calculation is first performed with traditional CC theory, producing a cluster operator $T_0 \in \mathcal{C}$ (and also a $\Lambda_0 \in \tilde{\mathcal{C}}$) such that $\psi_0 = e^{T_0}\phi$ is an approximate ground state. Using the theory of linear response, excited-state energies are approximated by the eigenvalues of a ``dressed'', or effective Hamiltonian, $A = P e^{-T_0}He^{T_0} P$, where $P$ is the orthogonal projector onto the configuration-interaction space, usually at the same level of truncation $\mathcal{T}$ as the underlying CC calculation. Thus, a bivariate Rayleigh quotient is set up, $\mathcal{E}(L,R) = \braket{\tilde{\phi}|L A R\phi}/\braket{\tilde{\phi}|LR|\phi}$, where $(L,R) \in \tilde{\mathcal{C}}\oplus\mathcal{C}$. The left and right eigenvectors of $A$ are treated as approximations to excited states. The action of TD-EOM-CC theory is simply
\begin{equation}
    \mathcal{A}_\text{EOM-CC} = \int i \braket{\tilde{\phi}|L\dot{R}|\phi} - \braket{\tilde{\phi}|LAR|\phi} \; dt,
\end{equation}
producing linear canonical equations of motion and a simple Poisson bracket, formally identical to Eq.~\eqref{eq:poisson-bracket-exact}.

\section{Conclusion}
\label{sec:conclusion}

In this article, we studied the time-dependent bivariational principle, and employed a differential geometric point of view. We introduced an action principle $\delta\calA = 0$, where the field variables are the wavefunction and its complex conjugate $(\tilde{\psi},\psi)$. Approximate propagation techniques of bivariational type are then obtained by restricting these to lie in a smooth submanifold, or ansatz space. 

We demonstrated that taking the real and imaginary parts $\Re\calA$ and $\Im\calA$ resulted in two independent variational principles that both reproduce exact dynamics when no approximations in the wavefunctions are introduced. A distinction was further made of approximate methods depending on the ansatz space being parameterized with complex or real coordinates. When complex coordinates are used, all variational principles are equivalent. When real coordinates are used, the real and imaginary principles are not always equivalent. Comparison with the time-dependent (uni-)variational principle and the McLachlan variational principle were made.

The imaginary principle is valid, yet its physical meaning is presently unclear, since the generator for time evolution is not the real part of the energy, but instead the imaginary part. It is not from the outset ``unphysical'', since bivariational methods invariable introduce formally compled-valued energies and expectation values. The imaginary principle ensures that the imaginary part of the energy is conserved and thus guaranteed to remain small. We relegate the study of this principle to future investigations.

In analogy with classical mechanics, Poisson brackets were introduced that allow analogy with the transition from classical to quantum mechanics. In particular, time evolution of observables become Poisson brackets.

In the final section of the article, we formulated various methods for real-time propagation using the TD-BIVP. We formulated time-dependent traditional coupled-cluster theory, orbital-adaptice time-dependent coupled-cluster theory and orthogonal orbital-optimized coupled-cluster theory, where the single-particle functions are allowed to move during dynamics, and time-dependent equation-of-motion coupled cluster theory.

\section{Acknowledgments}
This work has received funding from the Research Council of Norway (RCN) under
CoE Grant Nos.~287906 and 262695 (Hylleraas Centre for Quantum Molecular Sciences), and Independent Research Fund Denmark through Grant No. 1026-00122B.

\section*{Appendix}
\label{sec:appendix}

\subsection*{Time-independent bivariational principle}

In this section, let $u = (\tilde{\psi},\psi)\in\HH$. Consider initially the stationary bivariational principle: Let $A \in B(\calH)$ be a (not necessarily self-adjoint) operator,  let $\calN = \{ u \in \HH \mid \braket{\tilde{\psi}|\psi} = 0 \}$, and consider the bivariate Rayleigh quotient:
\begin{equation}
    \calE_A : \HH \setminus \calN \to \CC, \quad (\tilde{\psi},\psi) \mapsto \frac{\braket{\tilde{\psi}|A\psi}}{\braket{\tilde{\psi}|\psi}}.
\end{equation}
The functional $\calE_A$ is a bivariational expectation value functional for the operator $A$. It is a Fréchet smooth function, which means that it can be Taylor expanded in its argument about any point. Infinitesimal variations $\delta \calE_A = \dualpairing{\delta u,d\calE_A(u)}$ are expressed as directional derivatives, and $\delta\calE_A=0$ for all variations $\delta u$ if and only if $d\calE_A(u)=0$, if and only if
\begin{equation}
    \braket{\tilde{\psi}|\psi} \neq 0, \quad H\psi = a \psi, \quad A^t \tilde{\psi} = a \tilde{\psi}, \quad a = \mathcal{E}_A(\tilde{\psi},\psi).
    \label{eq:bivp-critical-point}
\end{equation}
Here, $A^t\tilde{\psi}$ can be written using bra notation as $\bra{\tilde{\psi}}A$. Thus, \emph{both} left and right eigenvectors for the eigenvalue $a$ must exist for a critical point to exist. This clearly restricts the class of non-selfadjoint operators that can be treated, but the main point is that $A$ being Hermitian was \emph{not} used in the proof of the bivariational principle, in contrast to the corresponding proof for the standard Rayleigh--Ritz variational principle. Instead, we are given the choice to approximate $\tilde{\psi}$ and $\psi$ with \emph{independent} approximations.

We observe that $\calE_A$ is invariant under individual phase and normalization transformations on $(\tilde{\psi},\psi)$, that is $\calE_A(\alpha\tilde{\psi},\beta\psi) = \calE_A(\tilde{\psi},\psi)$, for any $\alpha,\beta\in\CC$. Indeed, we can express $\calE_A = \Tr(A\rho)$, where $\rho = \braket{\tilde{\psi}|\psi}^{-1}\ket{\psi}\bra{\tilde{\psi}}$, which can be thought of as a pure state in the bivariational setting. Thus, we want \emph{all physical predictions} to be stated in terms of $\rho$.

The scale invariance of $\rho$ implies that the critical point condition (\ref{eq:bivp-critical-point}) does not yield locally unique solutions: For every critical point $(\tilde{\psi},\psi)$, another critical point is given by $(\alpha\tilde{\psi},\beta\psi)$ for any nonzero $\alpha,\beta\in\CC$. Moreover, if $a$ is not a simple eigenvalue (meaning that $a$ is a degenerate eigenvalue), there will be further degrees of freedom in the set of critical points.

\subsection*{The time-dependent bivariational principle in terms of the density operator}

The development in the present section is similar to the treatment of the time-dependent variational principle in Ref.~\onlinecite{kramer_geometry_1981}.
Consider the integral
\begin{multline}
    \mathcal{L} = \int_0^T \frac{1}{\braket{\tilde{\psi}|\psi}}\left(\frac{i}{2} \braket{\tilde{\psi}|\dot{\psi}} - \frac{i}{2}\braket{\dot{\tilde{\psi}}|\psi}\right) \\- \mathcal{E}_A(\tilde{\psi},\psi)  \; dt.
\end{multline}
This functional is invariant (up to a total time derivative) under \emph{time-local} phase and normalization transformation of $\tilde{\psi}(t)$ and $\psi(t)$ individually, and hence the critical point histories (solutions of the Euler--Lagrange equations) are unique only up to such changes. On the other hand, the time-dependent state $\rho(t)$ is unique. Denote the integrand in $\mathcal{L}$ by $L(t) = Q(t) - \calE_A$. The Euler--Lagrange equations are equivalent to:
\begin{equation}
    \braket{\tilde{\psi}(t)|\psi(t)} \neq 0, \quad (i \partial_t - A)\psi(t) - L(t) \psi = 0,
\end{equation}
and
\begin{equation}
    (-i\partial_t - A^t)\tilde{\psi} - L(t)\tilde{\psi} = 0.
\end{equation}
For every solution, $u(t)$, new solutions are generated by time-local phase and normalization transformations $u'(t) = (\alpha(t)\tilde{\psi},\beta(t)\psi)$. These leave $\rho(t)$ invariant. One particular set of solutions is the ``canonical'' solutions to the time-deoendent Schrödinger equation and its dual,
\begin{multline}
    \braket{\tilde{\Psi}(t)|\Psi(t)}\neq 0, \quad    (i \partial_t - A) \Psi(t) = 0, \\ (-i\partial_t - A^t)\tilde{\Psi}(t) = 0.
\end{multline}
One can see that all other solutions can be mapped to a solution on this form via a particular local phase and normalization transformation. It follows that all solutions with the same initial condition up to phase and normalization are equivalent to the canonical solution. Thus, $\mathcal{L}$ is the action functional $\calA$ from Section \ref{sec:bivp} expressed in terms of the density operator $\rho(t)$. Indeed, the canonical solutions to $\delta \mathcal{L}=0$ are precisely the solutions to $\delta\calA=0$.

Consider now complex or real submanifold $\calM \subset \HH$. For $\delta\calL=0$ to be equivalent to $\delta\calA=0$, we have two sufficient conditions: (1) $\calM$ is overlap normalized, i.e., $\braket{\tilde{\psi}|\psi}=1$. Indeed, then $\calL$ and $\calA$ becomes identical functionals. (2) $\calM$ contains ``rays'', i.e., if $(\tilde{\psi},\psi) \in \calM$, then $(\alpha\tilde{\psi},\beta\psi)\in \calM$ for all $\alpha,\beta \in \CC$. Indeed, then $\calL$ is still invariant under time-local phase and normalization changes. However, we observe, that for such a manifold, we can select a submanifold on the form (1) that will generate the same solutions using the functional $\calA$.

\bibliography{refs.bib}

\end{document}